\documentstyle[prl,aps,twocolumn]{revtex}

\begin{document}

\pagestyle{myheadings}
\markboth{Masahito Ueda and Anthony J. Leggett}
{Masahito Ueda and Anthony J. Leggett}

\draft

\title{Macroscopic Quantum Tunneling of a Bose-Einstein Condensate\\
with Attractive Interaction}

\author{Masahito Ueda$^1$ and Anthony J. Leggett$^2$}
\address{$^1$Department of Physical Electronics, Hiroshima University,
Higashi-Hiroshima 739, Japan\\
$^2$Department of Physics, University of Illinois at
Urbana-Champaign, Urbana, IL61801-3080}
\date{\today}

\maketitle

\begin{abstract}
A Bose-Einstein condensate with attractive interaction can
be metastable if it is spatially confined and if the number
of condensate bosons $N_0$ is below a certain critical value
$N_{\rm c}$.
By applying a variational method and the instanton techinique
to the Gross-Pitaevskii energy functional, we find analytically
the frequency of the collective excitation and the rate of
macroscopic quantum tunneling (MQT).
We show that near the critical point the tunneling exponent 
vanishes according to $(1-N_0/N_c)^\frac{5}{4}$ and that MQT
can be a dominant decay mechanism of the condensate for $N_0$ 
very close to $N_{\rm c}$.
\end{abstract}

\pacs{03.75.Fi, 73.40.Gk, 32.80.Pj, 05.30.Jp}

\narrowtext
An explosively increasing amount of research is being carried 
out on the phenomenon of Bose-Einstein condensation (BEC) in 
gases of $^{87}$Rb \cite{Anderson}, $^7$Li \cite{Bradley},
and $^{23}$Na \cite{Davis} atoms.
Unlike the other two species, a uniform system of $^7$Li atoms
is usually believed not to form a stable BEC state \cite{Nozieres}
because the s-wave scattering length $a$ is negative and the 
attractive interaction between the atoms causes the condensate
to collapse upon itself.
When atoms are spatially confined, however, they acquire zero-point
energies which, under certain conditions, counterbalance the 
attractive interaction, thereby allowing a metastable condensate 
to form.

As the number of condensate bosons increases, the  attractive
interaction becomes strong and the energy barrier that prevents
the condensate from collasing becomes accordingly low.
Given a confining potential which determines the zero-point 
energies, there exists a critical number $N_c$ of condensate
bosons at which the energy barrier vanishes.
When the number of condensate bosons $N_0$ is slightly below 
$N_c$, the energy barrier will be so low that the condensate 
might undergo macroscopic quantum tunneling (MQT) to a dense 
state.
Kagan {\it et al.} \cite{Kagan} estimated the overlap integral
between the metastable condensate and the dense state to be 
proportional to 
$\exp\!\left(-\frac{3}{2}N_0\ln\frac{l_0}{L^*}\right)$,
where $l_0$ is the amplitude of zero-point oscillations of the 
trap, and $L^* \sim 3|a|N_0$.
Shuryak \cite{Shuryak} estimated the MQT rate to be 
proportional to 
$\exp\left[-0.57(N_{\rm c}-N_0)\right]$.
Stoof \cite{Stoof} wrote down a WKB formula for the MQT rate but 
did not explicitly evaluate it near the critical point.

In this Letter we use a variational method and the instanton 
technique to show that the tunneling exponent vanishes faster
than found in Refs.\ \cite{Kagan,Shuryak}, namely as 
$(1-N_0/N_{\rm c})^\frac{5}{4}$, as $N_0$ approaches $N_{\rm c}$.
By comparing the MQT rate with other possible decay mechanisms,
we argue that MQT can be a dominant decay mechanism of the 
condensate near the critical point at zero temperature, contrary 
to the conclusions of Refs. \cite{Kagan,Shuryak}.
Since the formulas we obtain contain no fitting parameters, 
they can be used as stringent tests of existence of BEC.

At sufficiently low temperatures, a condensate of weakly 
interacting bosons is described by a single wave function 
$\Psi({\bf r})$, and the interaction between them is 
described by the s-wave scattering length $a$.
The wave function is determined so as to minimize the 
Gross-Pitaevskii energy functional \cite{GP}
\begin{eqnarray}
E[\Psi]&=&\!\!\int \!d{\bf r}  \Psi^*({\bf r})\!
\left[-\frac{\hbar^2\nabla^2}{2m}\!+\!V({\bf r})\!+\!
\frac{2\pi\hbar^2a}{m}|\Psi({\bf r})|^2\!
\right]  
\Psi({\bf r}),
\nonumber \\
%& & \times 
\label{GP}
\end{eqnarray}
where $m$ is the atomic mass, $V({\bf r})$ is a confining potential, 
and $2\pi\hbar^2a|\Psi({\bf r})|^2/m$ is the local mean-field 
interaction energy per particle.
The wave function $\Psi$ is normalized so that $\int|\Psi|^2 d{\bf r}$ 
is equal to the number of condensed bosons $N_0$. 
We assume an axially symmetric confining potential:
$V({\bf r})=m(\omega_\bot^2 x^2+\omega_\bot^2 y^2+\omega_\|^2 z^2)/2$,
where $\omega_\bot$ and $\omega_\|$ are respectively the frequency of 
the radial confining potential and that of the axial one.

To evaluate $E[\Psi]$, we assume a Gaussian trial wave function
\cite{Baym}
\begin{eqnarray}
\Psi({\bf r})=\sqrt{\frac{N_0}{\pi^\frac{3}{2}d_\bot^2d_\|}} 
\exp\!\left(-\frac{x^2+y^2}{2d_\bot^2}-\frac{z^2}{2d_\|^2}\right),
\label{wavefunction}
\end{eqnarray}
where $d_\bot$ and $d_\|$ are variational parameters.
Substituting Eq.\ (\ref{wavefunction}) into Eq.\ (\ref{GP}), 
we obtain
\begin{eqnarray}
& & f(r_\bot,r_\|)
\equiv\frac{4E}{N_0\hbar(\omega_\bot^2\omega_\|)^\frac{1}{3}}
\nonumber \\
& & =2\lambda^{-\frac{1}{3}}(r_\bot^2+r_\bot^{-2})
+\lambda^\frac{2}{3}(r_\|^2+r_\|^{-2})-\gamma r_\bot^{-2}r_\|^{-1},
\label{f}
\end{eqnarray}
where 
$r_\bot\equiv d_\bot\sqrt{m\omega_\bot/\hbar}\equiv d_\bot/d_{\bot 0}$
and $r_\|\equiv d_\|\sqrt{m\omega_\|/\hbar}\equiv d_\|/d_{\| 0}$
are the radial and axial widths of the condensate normalized by
their noninteracting values;
$\lambda\equiv\omega_\|/\omega_\bot$ is the asymmetry parameter
of the confining potential, and
\begin{eqnarray}
\gamma\equiv \frac{4N_0}{\sqrt{2\pi}}
\frac{|a|}{(d_{\bot 0}^2d_{\| 0})^\frac{1}{3}}
\label{gamma}
\end{eqnarray}
is the dimensionless strength of interaction

For a metastable condensate to exist, the function 
$f(r_\bot,r_\|)$ must have a local minimum which is 
determined from conditions
\begin{eqnarray}
\frac{\partial f}{\partial r_\bot}
=0 
\rightarrow 
r_\|=\frac{\lambda^\frac{1}{3}\gamma}{2(1-r_\bot^4)},
\label{cond1} \\
\frac{\partial f}{\partial r_\|}
=0
\rightarrow 
r_\bot^2=\frac{\gamma r_\|}{2\lambda^\frac{2}{3}(1-r_\|^4)}.
\label{cond2}
\end{eqnarray}
The mestability of the condensate is determined from the 
curvatures of $f(r_\bot,r_\|)$ which are calculated to be
\begin{eqnarray}
\frac{\partial^2 f}
{\partial r_\bot^2}
=16\lambda^{-\frac{1}{3}}>0, \ \ \ 
\frac{\partial^2 f}
{\partial r_\|^2}
=2\lambda^\frac{2}{3}(3+r_\|^{-4})>0,
\label{cond4} \\
\frac{\partial^2 f}
{\partial r_\bot\partial r_\|}
=-2\gamma r_\|^{-2}\left[\frac{2\lambda^\frac{2}{3}(1-r_\|^4)}{\gamma r_\|}
\right]^\frac{3}{2}\leq 0.
\label{cond5} 
\end{eqnarray}
The condition for a metastable condensate to exist is that
the curvature of $f(r_\bot,r_\|)$ at the local minimum is 
positive in all directions, that is,
\begin{eqnarray}
\frac{\partial^2 f}{\partial r_\bot^2}
\frac{\partial^2 f}{\partial r_\|^2}
-
\left(\frac{\partial^2 f}{\partial r_\bot\partial r_\|}
\right)^2 > 0.
\label{cond6} 
\end{eqnarray}
It is clear from Eqs.\ (\ref{cond4}) and (\ref{cond5}) that 
the condition (\ref{cond6}) is always satisfied at the 
limit of weak interaction $\gamma\rightarrow 0$, but 
that it is violated at the limit of strong interaction 
$\gamma\rightarrow \infty$.
In between there must be a critical value of $\gamma$ such 
that the left-hand side of (\ref{cond6}) is zero. 
It is given by
\begin{eqnarray}
\gamma_{\rm c}=\frac{\lambda^\frac{5}{3}(1-r_\|^4)^3}
{r_\|^3(1+3r_\|^4)}.
\label{gamma_c}
\end{eqnarray}
For $\gamma>\gamma_{\rm c}$ the local minimum becomes a saddle point
and the condensate becomes unstable, collapsing into a dense state.
Substituting Eq.\ (\ref{gamma_c}) into Eqs.\ (\ref{cond2}) 
gives
\begin{eqnarray}
r_{\bot\rm c}=\frac{\lambda^\frac{1}{2}(1-r_{\|\rm c}^4)}
{\sqrt{2r_{\|\rm c}^2(1+3r_{\|\rm c}^4)}},
\label{r_bot}
\end{eqnarray}
where $r_{\|\rm c}$ denotes the value of $r_\|$ at the
critical point. Substituting Eqs.\ (\ref{gamma_c}) and 
(\ref{r_bot}) into Eq.\ (\ref{cond1}) gives
\begin{eqnarray} 
\lambda^2=\frac{4r_{\|\rm c}^4(1+3r_{\|\rm c}^4)^2}
{(1-r_{\|\rm c}^4)^3(3+5r_{\|\rm c}^4)}.
\label{r_para}
\end{eqnarray}
We may use Eq.\ (\ref{r_para}) to simplify Eq.\ (\ref{gamma_c}) somewhat:
\begin{eqnarray}
\gamma_{\rm c}=\frac{4r_{\|\rm c}(1+3r_{\|\rm c}^4)}
{\lambda^\frac{1}{3}(3+5r_{\|\rm c}^4)}.
\label{gamma_c2}
\end{eqnarray}
For a given asymmetry parameter 
$\lambda\equiv\omega_\|/\omega_\bot$, a real positive root of 
Eq.\ (\ref{r_para}) gives $r_{\|\rm c}$. Substituting this into Eqs.\
(\ref{r_bot}) and (\ref{gamma_c2}) gives $r_{\bot\rm c}$ and 
$\gamma_{\rm c}$, respectively.

For an isotropic case ($\lambda=1$), we obtain
$r_{\bot\rm c}=r_{\|\rm c}=5^{-\frac{1}{4}}\simeq 0.669$ and
$\gamma_{\rm c}\simeq 1.07$ in agreement with the results of 
Refs.\ \cite{Fetter,Stoof}.
Taking experimental data from the second reference in \cite{Bradley}, 
where $a=-14.5$\AA, $d_{\bot 0}\simeq 3.08 \mu m$ and 
$\lambda\simeq 0.867$,
we obtain from Eq.\ (\ref{gamma}) that $N_{\rm c}\simeq 1460$.
This is 17\% greater than the more precise value of 1250 which is 
obtained by numerically solving the nonlinear Schr\"{o}dinger 
equation \cite{Ruprecht}. 
In evaluating the MQT rate, we will use the latter value for the 
critical number of condensate bosons.

\begin{figure}
\caption{
Critical values of the normalized radii 
$r_{\bot\rm c}\equiv d_{\bot\rm c}/d_{\bot 0}$, 
$r_{\|\rm c}\equiv d_{\|\rm c}/d_{\| 0}$, 
and the dimensionless strength of interaction $\gamma_{\rm c}$ 
as a function of the asymmetry parameter 
$\lambda\equiv\omega_\|/\omega_\bot$,
where $d_{\bot 0}\equiv(\hbar/m\omega_\bot)^\frac{1}{2}$ and 
$d_{\| 0}\equiv(\hbar/m\omega_\|)^\frac{1}{2}$.
Also plotted is the ratio $d_{\bot\rm c}/d_{\|\rm c}$ of the width 
of the condensate along the radial direction to that along the axial 
one. Note that 
$r_{\bot\rm c}$, $r_{\|\rm c}$, and $\gamma_{\rm c}$ 
refer to the left scale, while  
$d_{\bot\rm c}/d_{\|\rm c}$ refers to the right one.
}
\label{Fig1}
\end{figure}
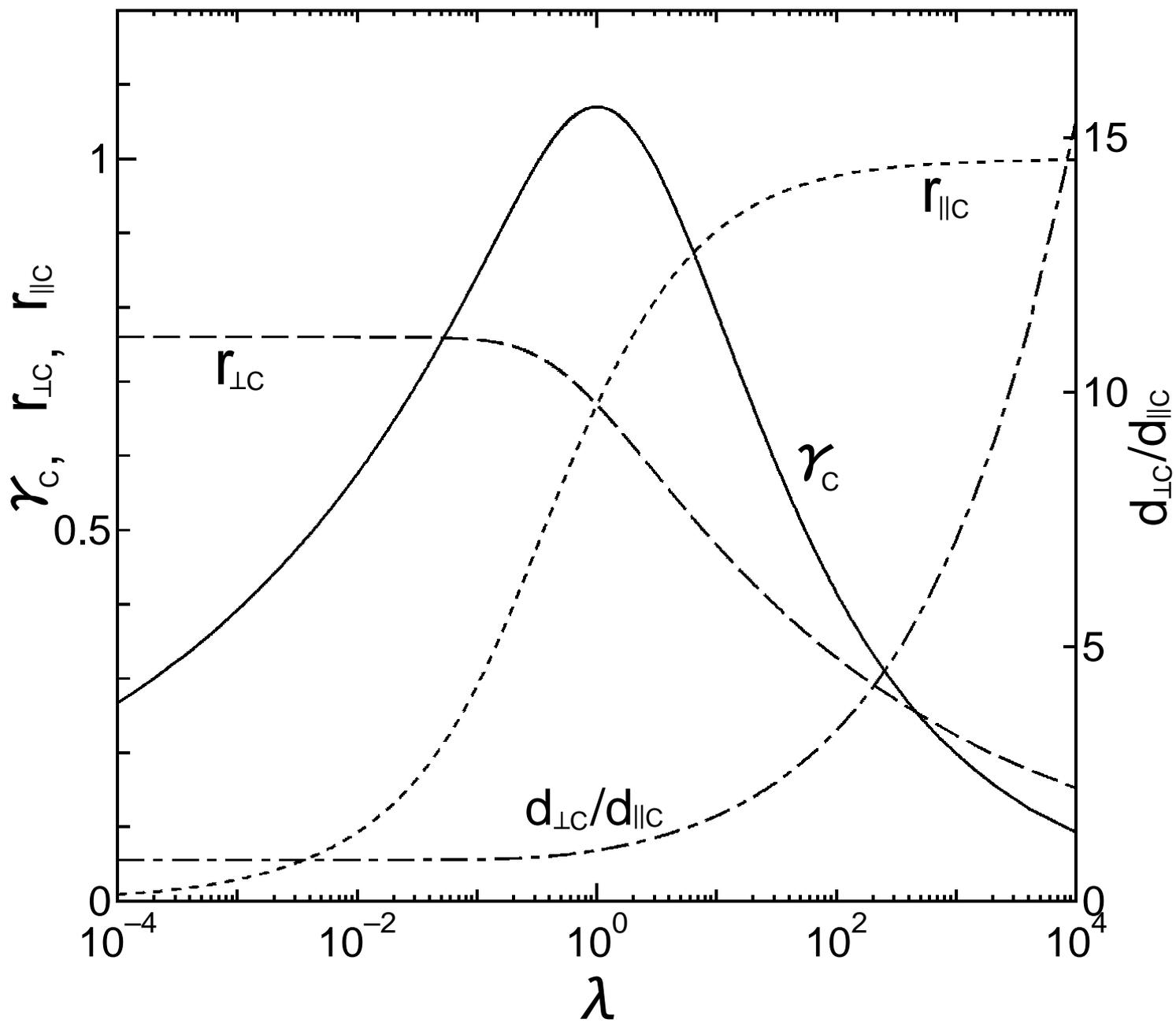

Figure \ref{Fig1} shows $r_{\bot\rm c}$, $r_{\|\rm c}$, 
and $\gamma_c$ as a function of the asymmetry parameter 
$\lambda$.
We see that the maximum value of $\gamma_{\rm c}$ can be attained
for the case of an isotropic potential \cite{Note}.
Also plotted is the ratio $d_{\bot\rm c}/d_{\|\rm c}$ of the width of
the condensate along the radial direction to that along the axial 
one. We note that the ratio remains relatively constant for 
$\lambda<1$ but it grows rapidly for $\lambda>1$.
In what follows we will focus on the case of an isotropic confining 
potential, and drop the subsrcripts $\bot$ and $\|$.
The function $f$ can then be written as
$f(r)=3r^{-2}+3r^2-\gamma r^{-3}$.

\bigskip
\noindent
{\it Collective excitation of the condensate}

\medskip
The condensate undergoes density oscillations around the local 
minimum.
Associated with this collective motion is a kinetic energy $T$
which we may write as
\begin{eqnarray}
T=3\zeta N_0 m\dot d^2=3\zeta N_0 m d_0^2\dot r^2
=3\frac{\zeta N_0\hbar}{\omega}\dot r^2,
\label{kinetic}
\end{eqnarray}
where $\zeta$ is a constant to be determined below.

The dynamics of the collective excitation is determined by the 
kinetic-energy term (\ref{kinetic}) and the quadratic part 
of the potential energy which is obtained by expanding 
$f(r)-f(r^{\rm min})$ in powers of $r-r^{\rm min}$:
\begin{eqnarray}
\frac{N_0\hbar\omega}{4}
\frac{f''(r^{\rm min})}{2}(r-r^{\rm min})^2,
\label{quadratic}
\end{eqnarray}
where $r^{\rm min}$ is determined from $f'(r^{\rm min})=0$.
For $\gamma$ slightly below $\gamma_c$, $r^{\rm min}$ is
close to $r_{\rm c}=5^{-\frac{1}{4}}$; that is, 
$r^{\rm min}=r_{\rm c}+\delta r$, where $\delta r$ is given
to leading order in $\gamma_{\rm c}-\gamma$ by
\begin{eqnarray}
\delta r\simeq\frac{r_{\rm c}^\frac{1}{2}}{2}
(\gamma_{\rm c}-\gamma)^\frac{1}{2}.
\label{deltar}
\end{eqnarray}
>From Eqs.\ (\ref{f}) and (\ref{deltar}), 
$f''(r_\bot^{\rm min})$ is calculated to be
\begin{eqnarray}
f''(r_\bot^{\rm min})\simeq
120\frac{\delta r}{r_{\bot c}}.
\label{2nd}
\end{eqnarray}
>From Eqs.\ (\ref{kinetic})-(\ref{2nd}) we obtain the frequency
$\omega_c$ of the collective mode near the critical point as
\begin{eqnarray}
\frac{\omega_c}{\omega}
=\sqrt{\frac{5}{\zeta}\frac{\delta r}{r_{\rm c}}}.
\label{omegac}
\end{eqnarray}

The constant $\zeta$ can be determined by a variational method. 
According to Ref. \cite{Bohigas}, an upper bound of the frequency of 
a collective mode is given by 
$\omega_c^{\rm upper}=\sqrt{m_3/m_1}/\hbar$, 
where 
$m_1=\langle 0|[F,[H,F]]|0\rangle/2$ 
and
$m_3=\langle 0|[[F,H],[H,[H,F]]]|0\rangle/2$ 
are the energy-weighted moment and the cubic-energy-weighted moment of 
the dynamic structure factor, with $F$ being an excitation operator.
For the monopole mode the excitation operator is given by 
$F=\sum_{i=1}^{i=N_0}(x_i^2+y_i^2+z_i^2)$.
Applying this formula to our Hamiltonian which is given in the square 
brackets of Eq. (\ref{GP}), we obtain Eq. (\ref{omegac}) with $\zeta=1/4$
and $\omega_c$ replaced by $\omega_c^{\rm upper}$.
Stringari \cite{Stringari} has pointed out that for the case of repulsive
interaction this method gives results in excellent agreement
with numerically obtained exact frequencies.
We expect that the same is true for the case of attractive interaction
and identify $\omega_c^{\rm upper}$ with $\omega_c$.
Substituting Eq.\ (\ref{deltar}) into Eq.\ (\ref{omegac}) and using 
Eq.\ (\ref{gamma}), we obtain
\begin{eqnarray}
\frac{\omega_c}{\omega}=
160^\frac{1}{4} \left(1-\frac{N_0}{N_{\rm c}}\right)^\frac{1}{4}.
\label{collective}
\end{eqnarray}
We thus find that as the number of condensate bosons $N_0$ approaches 
its critical value $N_{\rm c}$, the collective frequency $\omega_c$ 
vanishes as the one-fourth power of $1-N_0/N_{\rm c}$.

\bigskip
\noindent
{\it Rate of Macroscopic Quantum Tunneling}

\medskip
At sufficiently low temperature, the thermally activated decay of
the condensate over the barrier is negligible, and we can expect
MQT to provide a dominant decay mechanism.
When the barrier is low enough for MQT to occur but still so high
that the instanton approximation is valid, the MQT rate $\Gamma$ 
is given by $\Gamma=A e^{-S^{\rm B}/\hbar}$,
where $S^{\rm B}$ is the bounce exponent, that is, the value of the 
imaginary-time action $S$ evaluated along the bounce trajectory 
$r^{\rm B}(\tau)$.
Using Eq.\ (\ref{kinetic}) with $\zeta=1/4$ we obtain 
\begin{eqnarray}
S/\hbar&=&\frac{N_0}{4}\!\int\!d\tau\!
\left[3\dot r^2+f(r)-f(r^{\rm min})\right],
\label{action}
\end{eqnarray}
where the imaginary time has been rescaled as 
$\omega\tau\rightarrow \tau$.
The bounce trajectory $r^{\rm B}(\tau)$ is the one that makes 
$S$ extremal.
>From $\delta S/\delta r^{\rm B}=0$, we obtain an equation 
of motion for the bounce trajectory which can be integrated 
to give 
$(\dot r^{\rm B})^2=\frac{1}{3}[f(r^{\rm B})-f(r^{\rm min})]$.
Near the critical point, it is sufficient to expand the 
right-hand side of this equation in powers of 
$r^{\rm B}-r^{\rm min}$ and keep terms up to the third power:
\begin{eqnarray}
& & f(r^{\rm B})-f(r^{\rm min})
\nonumber \\
&\simeq&
\frac{f''(r^{\rm min})}{2}
\left[(r^{\rm B}-r^{\rm min})^2+
\frac{(r^{\rm B}-r^{\rm min})^3}{r^{\rm min}-r^{\rm L}}\right],
\label{expansion}
\end{eqnarray}
where 
$r^{\rm L}$ is the left turning point of $f(r)$ such that 
$f(r^{\rm L})=f(r^{\rm min})$.
The bounce trajectory is then obtained as
\begin{eqnarray}
r^{\rm B}(\tau)=
r^{\rm min}-\frac{r^{\rm min}-r^{\rm L}}
{\cosh^2 \left(
\sqrt{f''(r^{\rm min})/24}\tau\right)}
\label{btrajectory}
\end{eqnarray}
Substituting Eqs.\ (\ref{expansion}) and (\ref{btrajectory}) into 
Eq.\ (\ref{action}), we obtain
\begin{eqnarray}
S^{\rm B}/\hbar=\frac{2}{15}N_0
\sqrt{6f''(r^{\rm min})}
(r^{\rm min}-r^{\rm L})^2.
\label{B1}
\end{eqnarray}
Since 
$r^{\rm min}-r^{\rm L}
=3f''(r^{\rm min})/f'''(r^{\rm min})$,
we find from Eq.\ (\ref{2nd}) and 
$f'''(r^{\rm min})\simeq 120/r_{\rm c}$ that
$r^{\rm min}-r^{\rm L}=3\delta r$.
Hence
\begin{eqnarray}
r^{\rm min}-r^{\rm L}\simeq 
\frac{3}{2}\sqrt{r_{\rm c}(\gamma_{\rm c}-\gamma)}.
\label{difference}
\end{eqnarray}
Substituting Eqs.\ (\ref{2nd}) and (\ref{difference}) into 
Eq.\ (\ref{B1}), we obtain
\begin{eqnarray}
S^{\rm B}/\hbar&\simeq&4.58
N_0\left(1-\frac{N_0}{N_{\rm c}}\right)^\frac{5}{4}.
\label{bexponent}
\end{eqnarray}
Thus if the condensate is formed and its decay is governed by MQT,
the bounce exponent should be proportional to the five-fourths power 
of $1-N_0/N_{\rm c}$, in contrast to the unit power found 
in Ref. \cite{Shuryak}.

For a quadratic-plus-cubic potential, the prefactor $A$ is given by 
$A=\omega_c(15S^{\rm B}/2\pi\hbar)^\frac{1}{2}$
\cite{Caldeira}.
Substituting Eqs.\ (\ref{collective}) and (\ref{bexponent}) into this, we 
obtain
\begin{eqnarray}
\frac{A}{\omega}\simeq 11.8
N_0^\frac{1}{2}\left(1-\frac{N_0}{N_{\rm c}}\right)^\frac{7}{8}.
\label{prefactor}
\end{eqnarray}
Thus as $N_0$ approaches $N_{\rm c}$, the prefactor vanishes
as the seven-eights power of $1-N_0/N_{\rm c}$.
The crucial observation here is that the decrease in the prefactor
as $N_0\rightarrow N_{\rm c}$ is outweighed by the much faster 
increase in the exponential factor.
Because of this rapid growth in the MQT rate, MQT can be a 
dominant decay mechanism of the condensate near the critical 
point for experiments by the Rice group \cite{Bradley} as we 
now show.

Taking the experimental data of the second paper of 
Ref.\ \cite{Bradley},
we have $\omega=(\omega_x\omega_y\omega_z)^\frac{1}{3}
\simeq 908.4$/sec,
$\omega_c\simeq 3231(1-N_0/N_{\rm c})^{1/4}$,
$S^{\rm B}/\hbar\simeq 4.58 N_0(1-N_0/N_{\rm c})^{5/4}$, and
$A\simeq 10720 N_0^{1/2}(1-N_0/N_{\rm c})^{7/8}$ with $N_{\rm c}=1250$.
For $1-N_0/N_{\rm c}=10^{-2}$, we obtain 
$\omega_c\simeq 1022$/sec,
$A\simeq 6707$/sec,
$S^{\rm B}/\hbar\simeq 17.9$, 
and 
$\Gamma\simeq 1.12\times 10^{-4}$/sec.
For this $N_0$ MQT is negligible, but if $N_0$ is a little bit
closer to $N_{\rm c}$, e.g., for $1-N_0/N_{\rm c}=5\times 10^{-3}$, 
we obtain 
$\omega_c\simeq 859$/sec,
$A\simeq 3666$/sec,
$S^{\rm B}/\hbar\simeq 7.58$, 
and
$\Gamma=1.88$/sec, 
and the MQT rate is therefore significant.

It has sometimes been argued that the decay rate of the 
condensate due to MQT is much slower than that due to 
two-body dipolar and three-body collisions, and is 
therefore unlikely \cite{Shi}.
The above numerical evaluation, however, shows that near 
the critical point the MQT rate is at least comparable 
to the decay rate due to those inelastic collisions 
evaluated in Ref.\ \cite{Dodd,Dalfovo} because near the critical
point the MQT rate grows at an enormous rate as numerically
illustrated above.
Kagan {\it et al.} pointed out yet another interesting
decay mechanism due to exchange interaction \cite{Kagan}.
This mechanism becomes important when the mean-field interaction 
energy per particle is larger than the single-particle energy-level
spacing. For the experiments of Ref.\ \cite{Bradley}, 
these energies are estimated to be 1nK and 7nK, respectively, so 
the condensate is not likely to decay via this mechanism.

In conclusion, we have used a variational method and the instanton
technique to find analytically the frequency of the collective 
mode and the MQT rate of a Bose condensate with attractive 
interaction near the critical point of collapse.
The maximum number of condensate bosons is found to be attained
for the case of isotropic confining potential.
By comparing MQT with other decay mechanisms, we have argued that 
MQT can be a dominant decay mechanism of the condensate for $N_0$
very close to $N_c$.
The obtained formulas contain no fitting parameters and can 
therefore be used as rather stringent tests for the existence of 
BEC.
The fact that we have demonstrated the tunneling of the condensate 
to indefinitely low energy to occur does not necessarily imply that 
the physical system actually collapses, because we have not taken 
into account any effect of higher-order interactions.

We would like to thank R. Hulet and C. Sackett for valuable discussions 
and for sending us Ref. \cite{Note} prior to publication.
M. U. gratefully acknowledges helpful discussions with G. Baym
and I. Kostzin.
This work was supported by the Core Research for Evolutional
Science and Technology (CREST) of the Japan Science and Technology
Corporation (JST) and by the National Science Foundation under 
grant number DMR96-14133.

%%%%%%%%%%%%%%%%%%%%%%%%%%%%%%%%%%%%%%%%%%%%%%%%%%%%%%%%%%%%%%%%%%

\end{document}